\documentclass[reprint,aps,pre
]{revtex4-2}

\usepackage{amssymb}
\usepackage{amsmath}
\usepackage{bm}
\usepackage{graphicx}
\usepackage[toc,page]{appendix}
\usepackage{float}
\usepackage[colorlinks=true,allcolors=blue]{hyperref}
\usepackage[dvipsnames]{xcolor}
\usepackage{enumerate}
\usepackage{orcidlink}

\newcommand{\utrian}{\mathbin{\rotatebox[origin=c]{0}{$\blacktriangle$}}}
\newcommand{\dtrian}{\mathbin{\rotatebox[origin=c]{180}{$\blacktriangle$}}}
\newcommand{\ltrian}{\hspace{0.1em}\mathbin{\rotatebox[origin=c]{90}{$\blacktriangle$}}\hspace{0.1em}}
\newcommand{\rtrian}{\hspace{0.1em}\mathbin{\rotatebox[origin=c]{-90}{$\blacktriangle$}}\hspace{0.1em}}

\newcommand{\open}{\scalebox{1.3}{$\circ$}}
\newcommand{\bound}{\scalebox{1.3}{$\bullet$}}
\newcommand{\blocked}{\raisebox{0.4ex}{\scalebox{0.7}{$\bm{\oslash}$}}}

\newcommand{\new}[1]{#1}

\begin{document}
\title{Designing complex behaviors using transition-based allosteric self-assembly}

\author{Jakob Metson\,\orcidlink{0009-0004-5966-6074}}
\email{jakob.metson@ds.mpg.de}
\affiliation{Max Planck Institute for Dynamics and Self-Organization (MPI-DS), 37077 G\"ottingen, Germany}
\date{\today}

\begin{abstract}
Allosteric interactions occur when binding at one part of a complex affects the interactions at another part. Allostery offers a high degree of control in multi-species processes, and these interactions play a crucial role in many biological and synthetic contexts. Leveraging allosteric principles in synthetic systems holds great potential for designing materials and systems that can autonomously adapt, reconfigure, or replicate. In this work we establish a basic allosteric model and develop an intuitive design process, which we demonstrate by constructing systems to exhibit four different complex behaviors: controlled fiber growth, shape-shifting, sorting, and self-replication. In order to verify that the systems evolve according to the pathways we have developed, we also calculate and measure key length and time scales. Our findings demonstrate that with minimal interaction rules, allosteric systems can be engineered to achieve sophisticated emergent behaviors, opening new avenues for the design of responsive and adaptive materials.
\end{abstract}
                           
\maketitle

\section{Introduction}
Biological systems exhibit remarkable complexity and precision in their functionality, largely due to intricate molecular interactions. Among these, allosteric regulation \cite{Liu2016PCB, Bhagavan2002MBE, Cui2008PS} is particularly notable for its ability to modulate the activity of one site on a molecule in response to changes at another, often distant, site 
\cite{Wodak2019S, Cooper1984EBJ}. This non-local influence is fundamental to processes such as enzyme activation \cite{Monod1965JoMB, Reiss2020B, England2018PNAS, Kovbasyuk2004CR}, assembly \cite{Caspar1980BJ, Packianathan2010JV, Dubanevics2021JRSI, Lazaro2016JPCB, Hamilton2024PL, Packianathan2010JV, Perez-Riba2018PTRSBBS}
and signaling \cite{Chebaro2013PCB, Nussinov2014JACS, Thirumalai2018PTRSBBS}, where precise control over interactions is crucial for maintaining functionality. Due to its ubiquitous presence in biological systems, allosteric regulation has also proved an important consideration in disease and drug development \cite{Nussinov2013C, Abdel-Magid2015AMCL, Cecchini2015N, Dokholyan2016CR}.
The concept of allostery, while long studied in biological systems \cite{Pauling1935PNAS, Monod1961CSHSQB}, has seen growing interest in synthetic systems as researchers aim to develop and imitate complex behaviors. While biological systems benefit from millions of years of evolutionary refinement, artificial systems must be carefully designed to replicate natural behaviors. Recent advances in computational modeling and nanotechnology have begun to make this a feasible goal  \cite{Takeuchi2001ACR, Kremer2013C-EJ, Schneider2016OBC, Rocks2017PNAS, Flechsig2017BJ, Ha2017SPSMaP, Huang2018L, Campos2019NC, Zhang2022SA, Plaper2024CD}.
Understanding how to effectively design artificial systems that exploit allostery to produce targeted behaviors remains a rich area for exploration. 

In this work, we establish a simple allosteric model and develop an intuitive design process, by describing signal-passing allostery as transitions between different binding states. Using this framework we explore the design of four different behaviors: controlled fiber growth, shape-shifting, sorting, and self-replication.
These behaviors have been chosen to demonstrate a variety of interesting and practical use cases.

The configurability of allosteric interactions naturally make them a powerful tool for the control of assembly. For example, through sequential assembly \cite{Zhang2017NC}, allostery can be used to improve assembly yield by preventing the formation of incorrectly assembled or kinetically trapped structures \cite{Evans2024JCP}. We demonstrate this by designing a system to assemble fibers in a controlled manner.

Shape-shifting in the context of self-assembly refers to the ability of materials or structures to change form autonomously or in response to stimuli \cite{Nguyen2011AN, Denkov2015N, Osat2023NN, Metson2025arXiv-b, Shen2024NN}. This allows for the reuse of components, enabling adaptation to different tasks at different times \cite{Nagarkar2021PNAS, Sarraf2023SR, Barabas2023SA}. Here we employ allosteric interactions to trigger the release of components from a structure, leading to reconfiguration.

Allosteric interactions can also be used to create autonomous decision-making. We demonstrate this by designing a Maxwell's demon \cite{bhl57349} that is able to ensure that a particle always ends up on one side of a partitioned system, acting as a molecular sorter \cite{1879N}. Experimental DNA nanotechnology systems have successfully demonstrated similar behaviors through cargo sorting \cite{Thubagere2017S} and re-configurable DNA containers \cite{Andersen2009N, Chang2024JACS}.
We show how sorting and containing systems can be designed using allosteric interactions.

Self-replication is a cornerstone of living matter and has been a long-standing hallmark in designing synthetic systems. Indeed, intelligent artificial matter has its roots in research into self-replication \cite{Penrose1959SA, Neumann1966}, which has also been studied in the framework of signal-passing tile assembly \cite{Keenan2013DCMP, Alseth2024NC-1} and in other models \cite{Lano2025QAM, Zeravcic2014PNAS, Freitas2004, He2017NM}. We design a self-replicating system of dimers and investigate the resulting growth dynamics.

The designs we present are realistic enough to be feasibly built and each behavior is readily extensible to create more advanced functionality. For each process we analyze key time and length scales involved, allowing us to extract insights into the pathways driving these behaviors.
This work deepens our understanding of how allosteric principles can be applied to create complex, emergent behaviors in synthetic systems.

\section{Allosteric model}
To explore allosteric effects we establish a model containing basic allosteric interactions.
The model we consider is similar in concept to the signal-passing tile-assembly model \cite{Padilla2013UCNC, Cantu20203ISACI2}, which has its foundations in algorithmic assembly models \cite{Winfree1998, Rothemund2001, Lathrop2007CLRW}.

\begin{figure}[tb]
\centering
\includegraphics[width=0.9\linewidth]{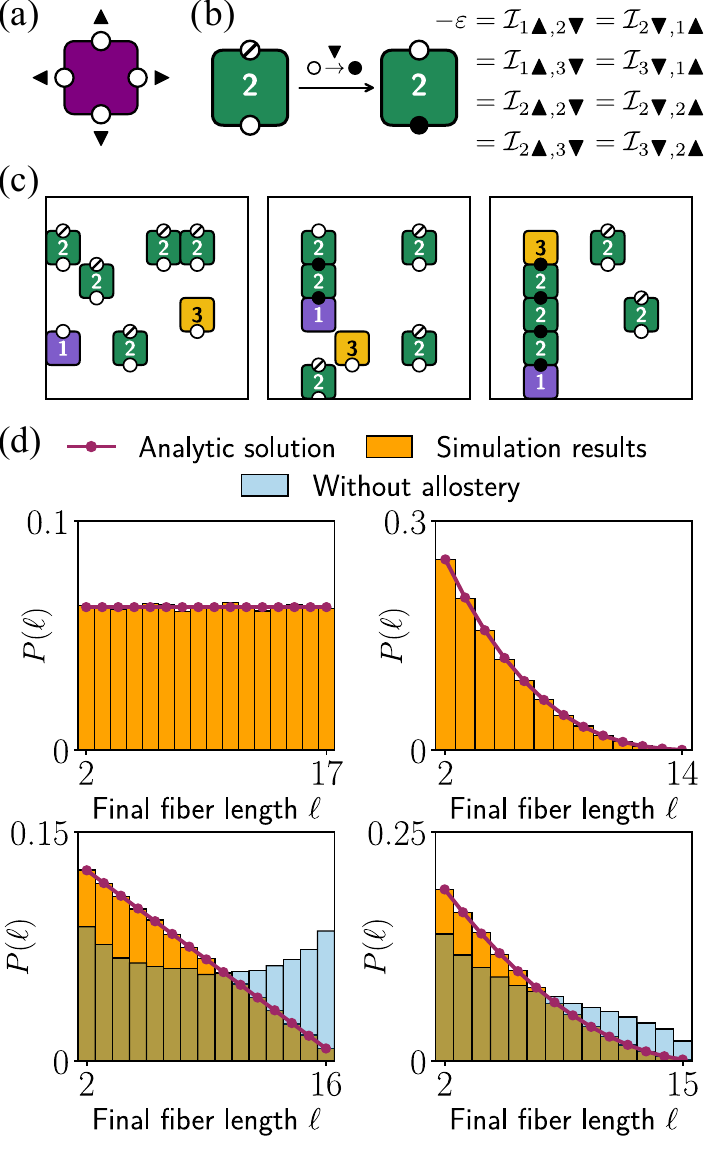}
\caption{Controlled fiber growth.
(a) Labeling of the binding sites.
(b) Allosteric transition and non-zero interaction matrix elements.
(c) Snapshots from a simulation of the fiber growth system.
(d) Distribution of final fiber lengths with \new{(anti-clockwise from top left) $N_3=1,2,3,4$}. Purple dots mark the analytic result Eq.~\ref{eq:final_length_distr}. Orange bars show the distributions resulting from $10^5$ simulations. \new{Light blue bars show the distribution resulting from $10^5$ simulations with the allosteric mechanism turned off.}
\label{fig:Fibres}}
\end{figure}

We consider two-dimensional systems consisting of a square lattice with periodic boundary conditions. A system is populated by individual units, which occupy a single site on the lattice and interact with their neighbors. When units bind, they form structures. Both structures and units are free to diffuse on the lattice, but cannot move into currently occupied positions.
Units on the lattice are not able to rotate. However, all of the behaviors designed here are robust to rotation since the binding sites can be made distinguishable, even with rotation. The main effect of including rotation would be to increase timescales since the units will not always be correctly aligned.
Each unit has four binding sites, labeled $\{\utrian,\dtrian,\ltrian,\rtrian\}$ as shown in Fig.~\ref{fig:Fibres}(a). The binding sites can take one of three states: unbound and available $\open$, bound $\bound$, or blocked $\blocked$.
The units belong to $n$ different species, labeled $s_i$, where $i$ ranges from $1$ to $n$. All units belonging to the same species have the same interactions and allosteric rules.  We use $N_i$ to denote the number of $s_i$ units in each system.
The matrix $\mathcal{I}$ describes the interactions between the different binding sites of each species. The matrix element $\mathcal{I}_{a\alpha,b\beta}=-\varepsilon$ if site $\alpha$ of species $a$ binds to site $\beta$ of species $b$. Otherwise, the matrix element is zero, meaning there is no interaction. We work in the tight binding limit where $\varepsilon \gg k_B T$. This means that once a bond has formed it can only be broken due to allosteric effects and not by spontaneous fluctuations.
We encode the allosteric effects using transition rules. The current state of a unit dictates which binding sites are blocked. Binding at a particular site can trigger the unit to transition to a new state, changing which sites are blocked. Take the allosteric transition shown in Fig.~\ref{fig:Fibres}(b) as an example. Initially the $\utrian$ site of $s_2$ units is blocked. Binding at the $\dtrian$ site acts as a trigger, changing the state of the $s_2$ unit and in this case releasing the blocking of the $\utrian$ site.
Using this framework, the design process simply consists of drawing the desired behavior and deducing the required states and transitions.

We use a Monte Carlo algorithm to simulate the system. At each timestep, a random unit or structure, and a random direction are chosen. If the desired positions to move into are currently occupied, the move is rejected and the system goes on to the next timestep. If the desired positions are unoccupied, then the unit or structure is moved by one lattice site in the chosen direction. Following a move, the interactions are recalculated based on the interaction matrix and any allosteric rules. This process repeats for the desired simulation time.

There has been compelling discussion and progress made in understanding the physics, such as entropic or energetic mechanisms, behind allostery \cite{Onan1983JACS, Hawkins2006BJ, Tsai2014PCB, LeVine2015E, Nussinov2015COiSB, McLeish2018PTRSBBS, Dutta2018PNAS, Vossel2023, Segers2023PRL}.
In this work we remain agnostic to the precise physical details, knowing that the \new{DNA strand-displacement} mechanism we have in mind is possible and has been demonstrated experimentally \cite{Padilla2015AC, Chatterjee2017NN}. We expand on the proposed allosteric mechanisms in the discussion section.
In the following sections we present the design of each desired behavior alongside analyses of the resulting dynamics.

\section{Applying the design process}
To illustrate the design process, we design four different target behaviors: controlled fiber growth, shape-shifting, sorting, and self-replication. The designs are presented in the following four subsections, and are discussed further in the discussion section.

\subsection{Controlled fiber growth}
In this system we use allosteric rules to enforce that free units only attach to growing fibers, preventing the formation of spurious sub-structures. 
\new{This results in precise control over the number of structures growing in the system.}
Fig.~\ref{fig:Fibres}(b) shows the interaction matrix and allosteric transition used for the fiber growth system. We use three species. First, the $s_1$ seed units act as a starting point for each fiber, with only the $\utrian$ site being active. The $s_2$ filler units can only bind above once they are bound below, preventing the formation of unwanted structures between free units. Finally, $s_3$ cap units only bind at the $\dtrian$ site. Once the fibers are capped off by an $s_3$ unit, they will no longer grow. This means that the fibers have a finite and well-defined final size.

In Fig.~\ref{fig:Fibres}(c) we show snapshots from a simulation of this system. When a filler unit reaches the square above the top of the fiber, it binds, growing the fiber. The binding of the filler unit at the $\dtrian$ site triggers a transition in the allosteric sequence, leading to the $\utrian$ site becoming available for binding. This allows for the further growth of the fiber. Over time, the fiber continues to grow via the addition of filler units, until a cap unit binds to the top of the fiber. This completes the fiber and the binding of the fiber will not change any more.

We use allostery to control how many fibers are grown (one per seed), by preventing the spontaneous nucleation of fibers. Another key quantity of interest is the length $\ell$ of the resulting fibers. Consider the growth of a single fiber in the system, with $N_1=1$. Since the units are initially randomly placed in the system and take random steps, on average any of the free units are equally likely to become the next addition to the fiber. Therefore, we can directly calculate an expression for the probability distribution of the system assembling a fiber of length $\ell$ as
\begin{equation} \label{eq:final_length_distr}
    P(\ell) = \frac{N_2! N_3}{(N_2+N_3)!} \; \frac{(N_2+2-\ell +[N_3-1])!}{(N_2+2-\ell)!},
\end{equation}
where $\ell$ takes integer values with $2\leq\ell\leq N_2+2$. For $N_3=1$ we get a uniform distribution and in general the distribution is a polynomial in $\ell$ of degree $N_3-1$. Figure~\ref{fig:Fibres}(d) shows this result plotted against simulations, with excellent agreement. This result applies generally, provided that each unit is equally likely to become the next addition to the fiber. It would not apply, for example, if the different species had different diffusion rates.
\new{The length distribution in a system with no allostery differs from the case with allostery, as shown in Fig.~\ref{fig:Fibres}(d). This demonstrates how allostery can be used to alter the length distribution itself, as well as controlling how many structures grow in the system.}
\new{An even tighter control over the growth process and length distribution could be obtained by introducing more species, which in combination with the allosteric interactions can be used to precisely manipulate which blocks get added and in what order.}

\begin{figure}[tb]
\centering
\includegraphics[width=0.9\linewidth]{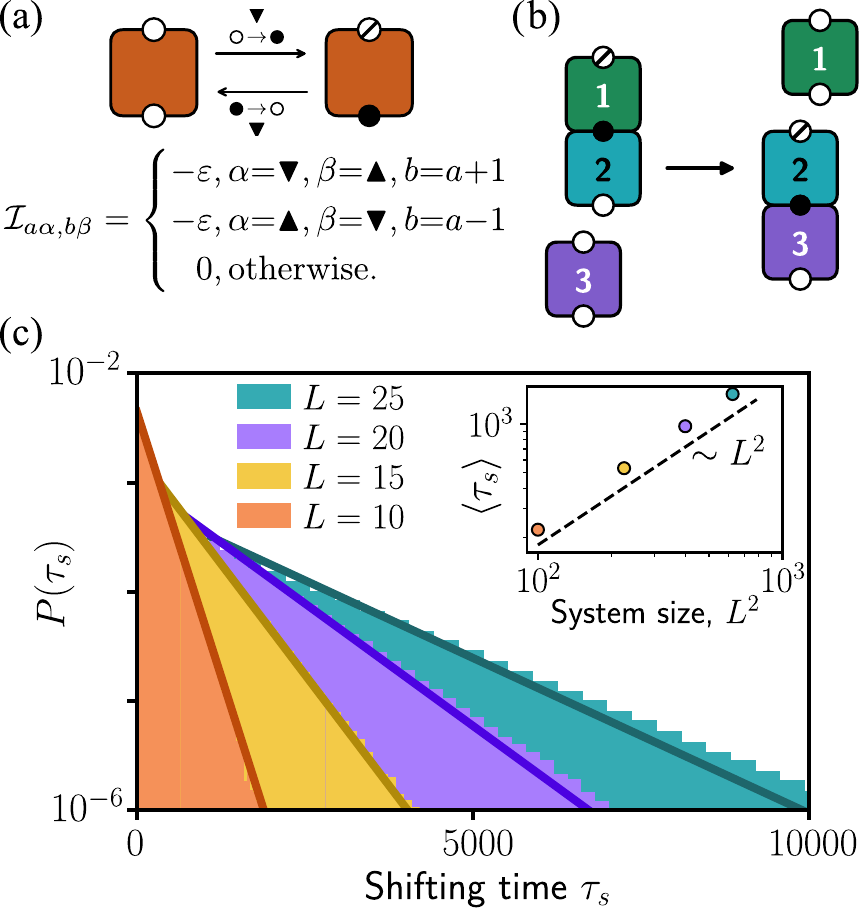}
\caption{Shape-shifting.
(a) Allosteric transitions and interaction matrix.
(b) Illustration of allosteric shape-shifting.
(c) Distributions of the shifting time $\tau_s$ on a semi-log plot, for different system sizes. The histogram uses $10^6$ simulation repeats for each system size. The solid lines plot exponential distributions. Inset: Scaling of the average shifting time with system size.
}
\label{fig:Shifting}
\end{figure}

\subsection{Shape-shifting}
To demonstrate shape-shifting we use a system of re-configurable dimers. We label the structures by the constituent species, such that a dimer formed from an $s_a$ unit above an $s_b$ unit is labeled $[a,b]$. 
Using $n$ species, we consider the $n-1$ dimers of the form
$D_i = [i,i+1]$, where $i$ ranges from $1$ to $n-1$.
We then aim to achieve the shifting sequence $D_1 \to D_2 \to \cdots \to D_{n-1}$.
The allosteric transitions, here applied to every species, and interaction matrix are shown in Fig.~\ref{fig:Shifting}(a).

The shifting process is sketched in Fig.~\ref{fig:Shifting}(b). We start with a $D_1$ dimer and a free $s_3$ unit. Once the $s_3$ unit reaches the position underneath $D_1$, it binds to the $s_2$ unit, to form structure $D_2$. The binding at the bottom of the $s_2$ unit triggers a shift in its allosteric sequence. This releases the bond between the $s_2$ and $s_1$ units, breaking apart the initially formed structure $D_1$ and leaving a complete $D_2$ structure.

The time for the structure to change, $\tau_s$, is given by the first passage time for the initially free $s_3$ unit to reach the lattice point underneath the structure $D_1$. Figure~\ref{fig:Shifting}(c) shows the probability distribution of $\tau_s$ measured in simulations. The solid lines are a plot of the exponential distribution $P(\tau_s) = \exp\left(-\tau_s/\langle \tau_s \rangle\right)/\langle \tau_s \rangle$,
where the average shifting time $\langle \tau_s \rangle$ is calculated using absorbing Markov chain techniques for the random walk \cite{Grinstead2006}. We find that $\langle \tau_s \rangle$ grows proportionally to the system size $L^2$, as shown in the inset plot of Fig.~\ref{fig:Shifting}(c).

\begin{figure}[tb]
\centering
\includegraphics[width=0.9\linewidth]{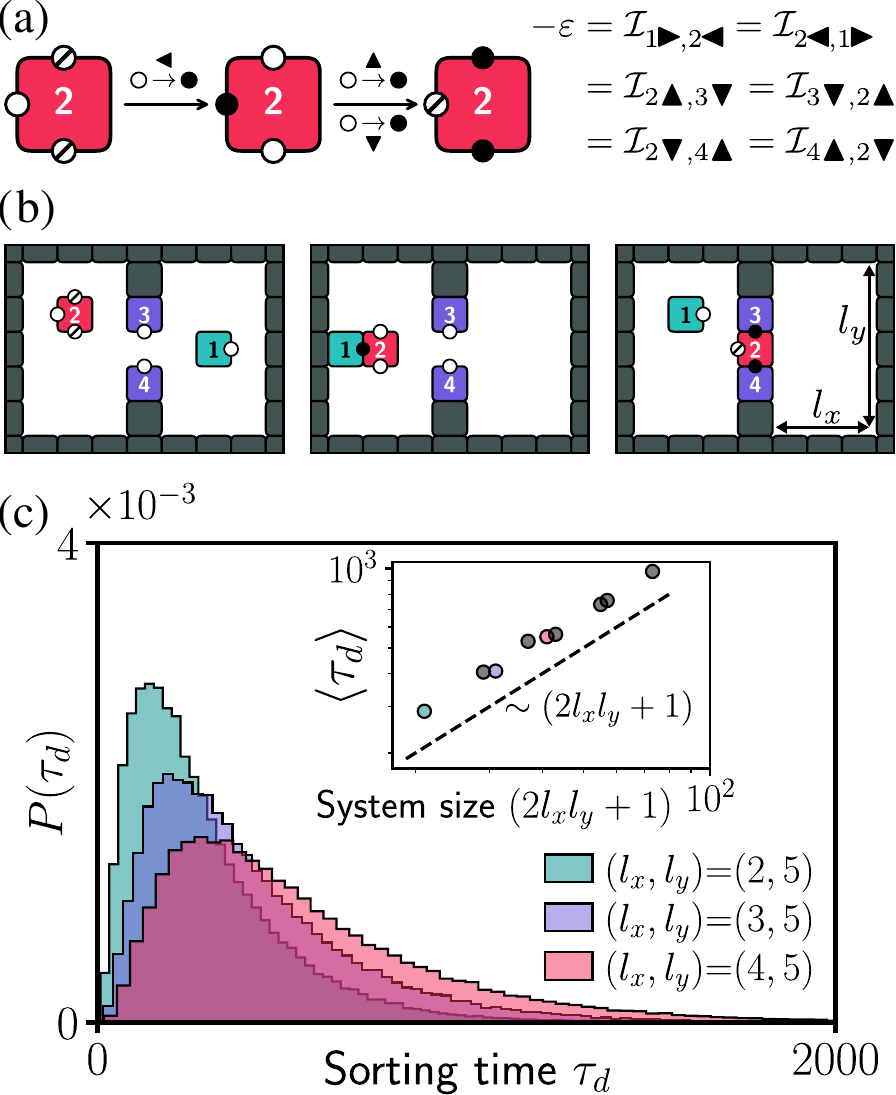}
\caption{Sorting. (a) Allosteric transitions and non-zero interaction matrix elements. (b) Snapshots of the sorting operation from a simulation. (c) Sorting time distribution for different system sizes. $(l_x,l_y)$ gives the dimensions of each half of the system. Inset: Scaling of the average sorting time with system size. For each system size we use $10^5$ simulations.
\label{fig:Demon}}
\end{figure}

\subsection{Sorting}
To set up the sorting system we use four species. The species of the target unit to be sorted is $s_1$. We use an $s_2$ unit as a demon trapdoor, which is able to seal off the two halves of the system, but only when the target unit is in the left half of the system. The $s_3$ and $s_4$ hinge species are immobile and act as anchoring points for the trapdoor. 
Only the demon trapdoor unit is allosteric, with the transitions and interaction matrix for this system shown in Fig.~\ref{fig:Demon}(a). We additionally use an immobile and non-interacting unlabeled species to create walls.

Figure~\ref{fig:Demon}(b) shows the operation of the demon. The target unit and demon trapdoor diffuse around the system until the trapdoor binds to the right-hand side of the target. This activates the binding sites at the top and bottom of the demon trapdoor. The complex diffuses until the trapdoor binds to the hinges, allosterically releasing the target particle and trapping it in the left-hand side of the system. Because of the binding geometry it is guaranteed that the unit to be sorted will always be released into the left half of the system, even in a system with rotation. Since the top and bottom binding sites of the trapdoor can be specified uniquely, if the target-trapdoor complex diffuses through the opening upside down, it will pass through without binding. The complex will only bind to the hinges when it has the correct orientation, guaranteeing that the target is released into the left-hand side of the system.

We are interested in the sorting time $\tau_d$, and how it depends on the system size. We initialize the system with a demon in the left half of the system and a target unit in the right half of the system, which represents an upper bound on the sorting time.
Figure~\ref{fig:Demon}(c) shows the distribution of sorting times for different system sizes. As the system size increases, the $\tau_d$ distribution broadens and the average sorting time $\langle\tau_d\rangle$ increases. In particular we find that the average sorting time scales linearly with system size, as shown in the inset plot of Fig.~\ref{fig:Demon}(c).

\new{Physically, in our mechanism-agnostic model, the change in entropy is compensated by the binding energy changes during the allosteric transitions.
In a DNA strand-displacement implementation, for example, the potential binding energy is stored in the DNA hairpin arrangements when the system is prepared.}

\begin{figure}[tb]
\centering
\includegraphics[width=0.9\linewidth]{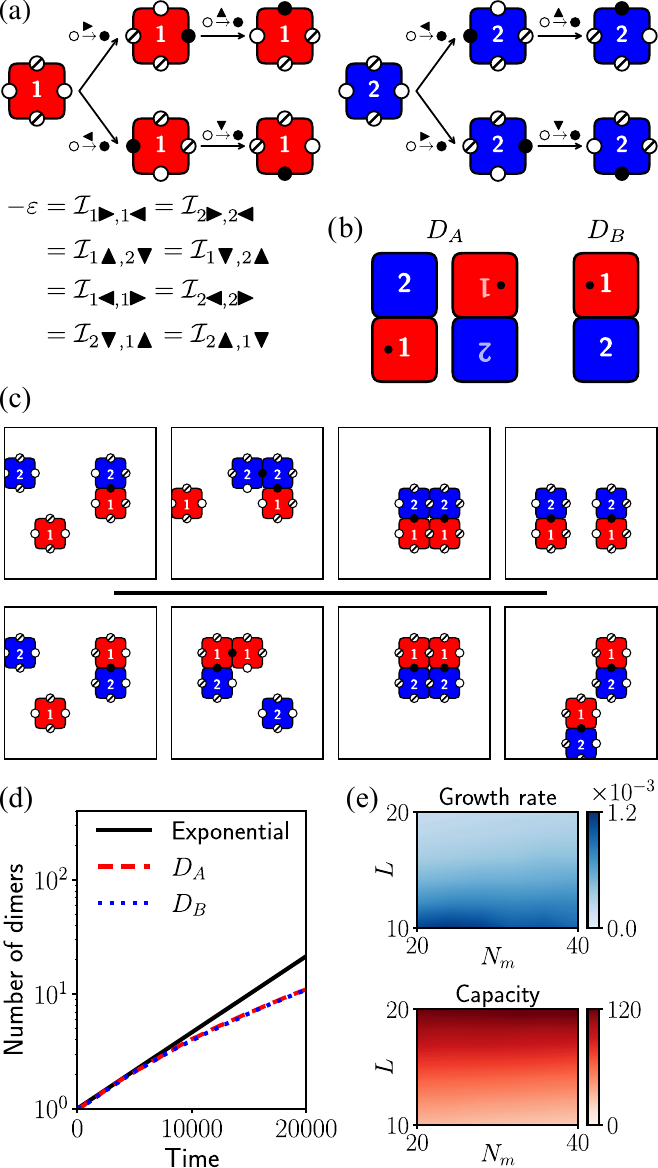}
\caption{Self-replication. (a) Allosteric transitions and non-zero interaction matrix elements. (b) The two dimers, $D_A$ and $D_B$. The small black dot is used to show that the dimers are mirror images of each other. (c) Snapshots from a simulation showing the self-replication of $D_A$ (top) and $D_B$ (bottom). (d) Number of dimers over time in a system initialized with one $D_A$ dimer and one $D_B$ dimer, on a semi-log plot.
Averaged over $10^3$ simulations. (e) Dependence of the growth rate and capacity on system size $L$ and number of monomers $N_m$.
\label{fig:Replicators}}
\end{figure}

\subsection{Self-replication}
To construct a self-replicating system we use two species, $s_1$ and $s_2$. The system forms two different dimer structures: $D_A$ with an $s_2$ unit above an $s_1$ unit and $D_B$ with an $s_1$ unit above an $s_2$ unit.
The dimer structures are able to produce identical copies by consuming unbound units floating in the bulk.
The interaction matrix and allosteric transitions for self-replication are given in Fig.~\ref{fig:Replicators}(a).
We furthermore impose the rule that new bonds can only be made between a structure and a free unit. 
\new{Using a DNA strand-displacement mechanism, this can be achieved by attaching free strands to the monomers, which bind to complementary strands on the structures. When a monomer binds to a structure, strands are released that are complementary to the free strands on other monomers. 
For the simulations, we model this through two unique horizontal binding sites to represent the strands on the monomers. 
This mechanism is reminiscent of the behavior explored in fiber growth, where a unit can only bind to other free units after binding to a structure.}

Since free units are consumed to form new structures, in these simulations we maintain a constant number of free units of each species to represent a bulk reservoir of monomers.
The two dimers, which we show in Fig.~\ref{fig:Replicators}(b), are unique - even in systems with rotation. To explicitly highlight this, we can place a spot on one side of the $s_1$ units. With the spot breaking the symmetry of the $s_1$ units, the two dimers are chiral mirror images of each other, which is illustrated in Fig.~\ref{fig:Replicators}(b) by rotating the $D_A$ dimer.

The self-replication process is demonstrated in Fig.~\ref{fig:Replicators}(c). One of the free units binds to the dimer, forming an intermediate structure. This intermediate structure continues to diffuse until the second free unit binds to it. At this point the horizontal bonds cleave, leading to the creation of a new dimer which is an identical copy of the original. The dimers only reproduce dimers of the same type. This can be seen in the two rows of snapshots in Fig.~\ref{fig:Replicators}(c). In the top row we initialize a system with a $D_A$ dimer and some free units, and a new $D_A$ dimer is created. Conversely, in the bottom row the initially present $D_B$ dimer creates a new $D_B$ dimer.

In Fig.~\ref{fig:Replicators}(d) we plot the number of dimers over time, for a system initialized with one $D_A$ dimer and one $D_B$ dimer, averaged over many realizations. Initially the growth is exponential since each new dimer created is able to form more dimers. However, over time the growth rate slows due to volume exclusion effects in the finite-sized system. Since neither dimer type has a competitive advantage, on average they grow at the same rate. 
To characterize the growth in more detail, we measure the dimers' growth law. By fitting the logistic growth equation $\dot{N}_{D} = rN_D(1-N_D/K)$, where $N_D$ is the number of dimers, we extract the growth rate $r$ and capacity $K$. The results are shown in Fig.~\ref{fig:Replicators}(e). As expected, the growth rate decreases with the system size $L$. In large systems, the growth rate increases with the number of monomers $N_m$, whilst in small systems increasing $N_m$ can actually reduce the growth rate due to volume exclusion becoming dominant. The capacity increases with $L$ and weakly decreases with $N_m$.

\section{Discussion}
Our transition-based design process allows for the rapid development of allostery-driven complex behaviors, such as those presented in this work. 

The assembly of fibers in the fiber growth system exhibits controlled growth. Due to the allosteric interactions, fibers only grow out from a seed, without the formation of other spurious structures or spontaneously nucleated fibers. The probability distribution for fiber lengths is well captured by the derived analytic expression, due to the predictability and control induced by the allosteric design. Alternative, but similar, mechanisms for the control of assembly have also been recently explored \cite{Bassani2024AN}, for example using non-reciprocal interactions \cite{Osat2024PRL, Navas2024TJoCP} or using interactions that can strengthen once correctly configured \cite{Zhu2024PRR}.

The shape-shifting concepts we present for simple dimer structures can readily be expanded to the shape-shifting of more complex structures, with allosteric interactions being used to coordinate the formation and release of different structures or sub-structures in a desired sequence.

By embedding decision-making rules directly into the system's interactions, we can construct a demon trapdoor that is able to decide when to close based solely on local information from its environment. This dynamic decision process is encoded in the allosteric transitions that govern how the trapdoor responds to different binding stimuli, allowing it to autonomously guide a particle into the desired compartment.
This suggests broader implications for the design of molecular machines and nanoscale systems, within which autonomous, self-regulating behaviors could be achieved by embedding smart rules at the molecular level.

The allostery-driven self-replication process means that the structures exclusively generate new structures of the same type, underscoring how allosteric interactions can be used to robustly develop non-trivial behaviors.

Recent proposals \cite{Evans2024JCP} have suggested DNA strand displacement mechanisms \cite{Simmel2019CR} to mediate the allosteric and signal-passing interactions in synthetic systems. 
In particular, DNA strand displacement mechanisms have been used to activate new binding sites by releasing strands. 
This mechanism could also be used to cleave existing bonds, by releasing a strand that displaces one of the strands in a bond between two different units.
An important step towards realizing this mechanism in practice would be to test its feasibility using detailed DNA simulations \cite{Poppleton2023JOSS} or by constructing experimental proofs of concept.
However, the model we employ in this work is general, and applies to any allosteric signal-passing mechanism.

\section{Conclusions}
The model presented here illustrates the versatility of allostery in driving complex dynamics in multi-species systems.
In this work we emphasize the use of allostery to perform practically meaningful tasks in self-assembled soft matter systems. 
Going beyond pure assembly, the integration of signal-passing decision-making with assembly processes leads to functional systems with autonomous dynamic functionality.
This includes the shape-shifting, sorting, and self-replication presented here.
Our design process presents new possibilities for the application of allosteric principles in the design of synthetic systems, particularly in the realms of nanotechnology and molecular robotics.

\nocite{AllostericData}

\acknowledgments

J.M. would like to thank Ramin Golestanian, Carl Goodrich, Saeed Osat, Illiya Stoev, and Petr Šulc for fruitful discussions, and Luca Cocconi and Navdeep Rana for insightful feedback. J.M. is grateful for the support of the Department of Living Matter Physics at the MPI-DS and the IMPRS-PBCS.


\bibliography{refs_JM, refs_Data}



\end{document}